# Structural Evolutions in Atoms of the Elements Executing Confined Interstate Electron Dynamics

**Mubarak Ali**

Department of Physics, COMSATS University Islamabad, Islamabad Campus, Park Road, 45550, Pakistan, mubarak74@mail.com, mubarak60@hotmail.com, http://orcid.org/0000-0003-1612-6014

**Abstract:** Differentiating structural evolution from structural development or formation opens many avenues of research. The study particularly advances the chemical and physical sciences, material science, energy science, and chemical engineering. By attaining uniform dynamics, atoms of suitable elements amalgamate. Atoms bind by executing confined interstate electron dynamics. Atoms execute electron dynamics in their original zones. For this purpose, atoms of suitable elements first attain a neutral state. The electrons of dynamics regain the state instantaneously upon the disappearance of the conservative forces. One cycle of the electron dynamics is sufficient to generate a binding energy. The shape of energy is similar to the trajectory of electron dynamics. The exerted forces remain almost in the actual formats of the growth of those atoms. Structures evolve into suitable gaseous element atoms above the ground surface, semisolid atoms at the ground surface, and solid atoms below the ground surface. The electrons executing dynamics simultaneously determine the structural dimension in atoms of different elements. Binding in gaseous atoms is from the upper side. The atoms in the solid elements bind from the downward sides. Both chemical force and energy bind nucleated mono-layers. The study also discusses a surface plasmon phenomenon. The structural evolution of atoms of suitable elements discussed here provides a new horizon for material and chemical science.

## 1.0. Introduction

In the study of a crystal structure, a unit cell defines the basis of the atoms. Seven crystal systems are well-known. In conventional studies, the patterns repeat at the points of Bravais Lattices. A Bravais lattice defines a three-dimensional space. There is no clue for the actual structure in different materials.

There is a structural development, formation, or evolution in different materials. They are different from the Bravais crystal systems. A structure of colloids, thin films, particles, composites, and nanostructures develops. It is due to a synthetic protocol. Here, a discussion is about the development mechanism.

Input parameters of the process largely control the developing structure of gold particles, as discussed elsewhere [1]. When a material resolution is down to a few nanometers, a microscopic study can reveal its process of structural development. It was not the issue with a developing structure. It was the issue of understanding and discussing it. An input power can control the dynamics of atoms.

The development of structures with different shapes is subject to synthetic protocols [1-5]. Atoms at work reveal charge dynamics, as observed in advanced microscopy [2]. From the amalgamation of particles in solution, it is possible to trace the force and energy behaviors [3]. It is also possible to observe the atomic structure of tiny particles from advanced microscopy [4]. The structure of platinum nanoparticles provides insight at the nanoscale, as discussed elsewhere [5].

In gallium arsenide nanowires, the crystal phase changes under various growth conditions [6]. Tuma *et al*. [7] discussed the physics of the phase transition of neurons from an amorphous state to a crystalline state. Zhao and Yang studied the structure of indium selenide by varying the pressure, as discussed elsewhere [8]. Rensberg *et al*. [9] demonstrated a phase transition in vanadium dioxide, and the optical properties changed depending on the attained state.

Studies based on gold particles [1, 3, 4, 10, 11], silver particles and binary composition particles [4], and carbon films [12, 13] have discussed the amalgamations of atoms. In developing a triangular-shaped tiny particle, the involved dynamics are discussed elsewhere [14]. A study elsewhere [15] discusses the modes of packing tiny-shaped particles in developing particles of high aspect ratios. Approximately 25 to 144 gold atoms developed a non-face-centered cubic geometric structure, as discussed elsewhere [16].

Many studies have also shown live visual images of amalgamating particles. Only some of the studies are cited here [17-21]. A photonic current is a source of processing materials in all methods and processes, rather than an electric current, as discussed elsewhere [22]. Understanding the difference between photons and electrons was also discussed [23]. Atoms of gaseous and solid elements deal with forces at the electron level, as discussed elsewhere [24]. In a graphite state, carbon atoms exhibit different behaviors [25]. In the synthesis of carbon films, the binding energy at different rates is discussed under varying chamber pressures [26].

The ground points of the gaseous atoms should remain above the ground or surface level. The binding in atoms of the gaseous elements should deal with the force of the space format. The atoms of the semisolid elements should keep the ground points at the ground or surface level. The semisolid atoms should bind under different conditions than solid atoms. Atoms should address the forces by remaining in their respective growth formats or zones.

The solid element atoms can also bind under different conditions. The ground points in solid atoms are much below ground or surface level. A hard coating is due to different ground points of solid and gaseous atoms [27]. The structure plays a central role in governing an application. It is a source of energy transportation. This study discusses the structural evolution in atoms of all suitable elements. The possibility of the surface plasmon phenomenon in a single-layer tiny particle is also part of this study.

## 2.0. Experimental details

The present study does not contain any experimental details. This study is suitable for the ongoing research activities related to material processing. This paper can support the studies related to improving the material design, simulation, and computation. In addition to the structural evolution of atoms of suitable elements, the current study also helps to understand energy phenomena at the electron level.

The results of the present study can be helpful wherever further research investigations address the structure of a material. A structure can be related to development or formation. This study also helps to understand light-matter interactions.

The study also covered areas such as energy science, nanoscience, surface science, and materials science.

In light of this work, several avenues of research are possible. Chemical science, materials science, energy science, nanoscience, and chemical engineering are core areas. The study also discusses the surface plasmon phenomenon in a tiny particle of a mono-layer. In atoms of all suitable elements, setting new experimental procedures for structural evolution is necessary.

The procedure for setup is simple and cost-effective for structural evolutions in atoms of semi-solid elements. Atoms in those elements already reside near the ground surface. The natural environment is already prevailing at a suitable level. The zones in gaseous and solid atoms remain above and below the ground surface. Thus, the setup in their structural evolutions is crucial and more effort-demanding.

## 3.0. Models and discussion

The force and energy of the process bind atoms for developing structures, as discussed elsewhere [11, 14]. In those studies, the development of the structures was owing to the synthetic protocols. Electrons of a carbon atom in any state do not have conserved forces [25]. Therefore, atoms of any carbon state do not <u>evolve</u> a structure. There is a <u>formation</u> of structure. A study elsewhere [27] identifies the role of a non-conservative energy and force in developing a hard coating.

The purpose of the study is to discuss the structural evolution in atoms of all suitable elements. The study does not discuss the development of a structure. In this study, the outer ring electrons of the atoms experience a conservative force. When an atom attains a neutral state for its under-study electron (an outer-ring electron), it can deal with inter-state electron dynamics.

A conservative force on the electron should be the cause of confined interstate dynamics. A conservative force is a non-frictional force. Forces from two poles track the dynamics of the under-study electron at a time. The forces exerted to turn the electron during its dynamics depend on the atomic structure.

The shape of the tracing trajectory due to the electron dynamics should depend on the distance or path between two states. The following sections also refer to both conservative force and polar force. A structural dimension depends on the number of outer-ring electrons of the atoms executing inter-state dynamics simultaneously. The forthcoming sections also target the study of structural dimensions to some extent.

### *3.1. Structural evolutions of gaseous element atoms*

The level of amalgamation of gaseous atoms above the ground or surface depends on the ground point of their gaseous element. The force above the ground or surface relates to a force in space format. A suitable electron of the outer ring first reaches the neutral state to evolve the structure, which can be just for an instant. In the original zone or format, the outer ring electrons of suitable element atoms can remain in their neutral states.

As the original zone of a gaseous atom is the space format or atmosphere, the main force exerted on that electron remains along the north pole. There is a minor contribution of the force from the east-west poles. A force from the east pole or the west pole contributes to the electron. Gaseous atoms evolve a structure in a gaseous, misty, or vaporous form.

Label (1) in Figure 1 denotes the orientation of the left-sided electrons of the gaseous atoms. Electrons maintain a ~ 5° orientation along the north pole. Label (2) in Figure 1 denotes the orientation of right-sided electrons. Electrons also maintain a ~ 5° orientation along the north pole. These orientations of the electrons are from the centers of their atoms. Label (3) in Figure 1 indicates the ground points of the atoms in a space format. A curved line in Figure 1 denotes the ground points of the atoms in space format. Atoms in the space format are the gaseous elements. In Figure 1, label (4) indicates the ground or surface level.

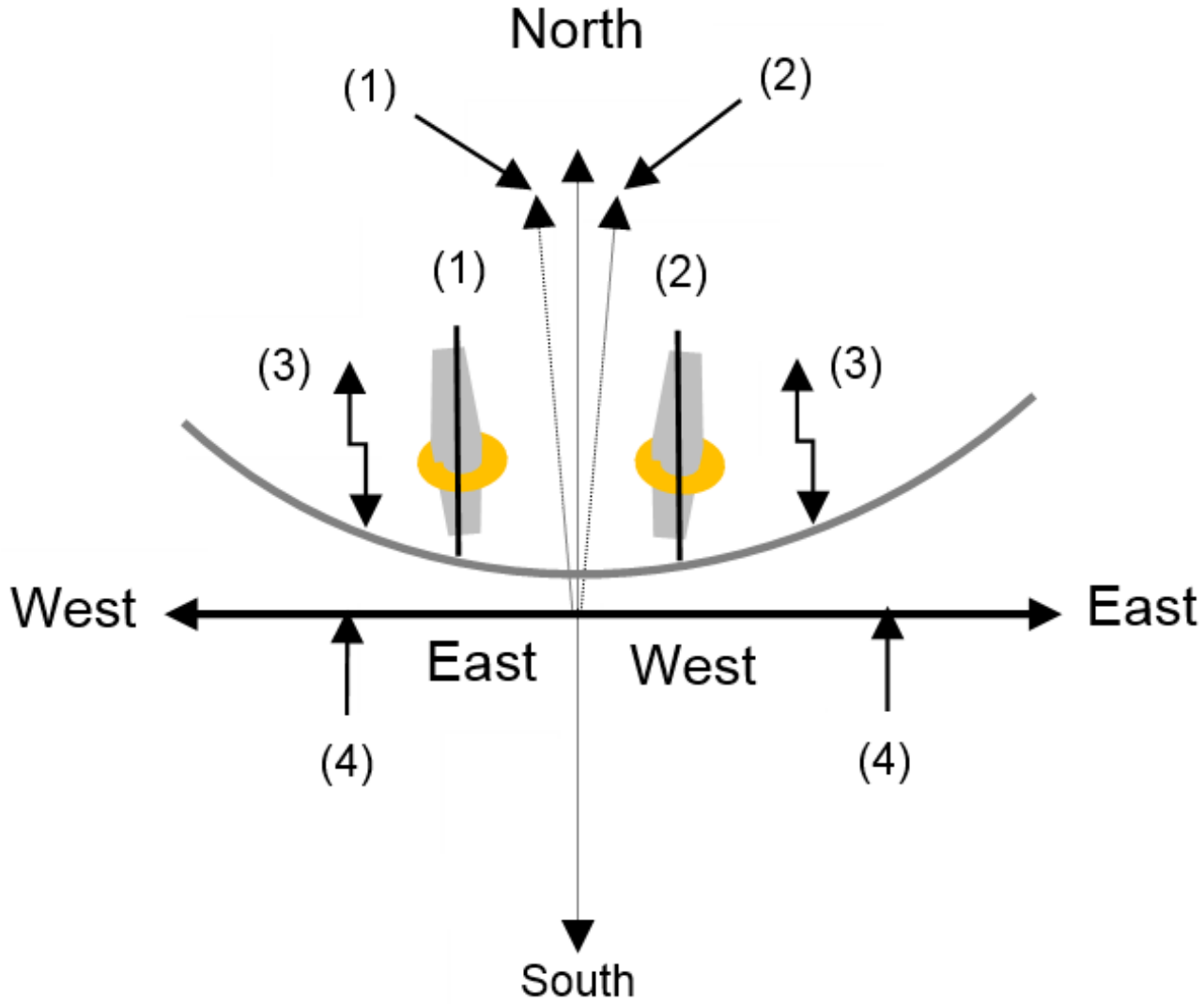

**Figure 1:** Exerted forces on the electrons of suitable gaseous element atoms to generate binding energy in one cycle of dynamics: (1) orientation of the electron ~ 5° left from the drawn vertical line, (2) orientation of the electron ~ 5° right from the drawn vertical line, (3) ground points of suitable gaseous element atoms and (4) ground or surface level. (Sketch drawn in estimation)

A suitable outer ring electron enters the neutral state. For this, the atom adjusts the energy knots to achieve a neutral state for that electron. That electron leaves the state as per the conservative force. At the electron level, a conserved energy is already there. The electron dynamics of one cycle generate binding energy. The electron does not connect to the occupied energy knot, as discussed elsewhere [23]. After the completion of one cycle of the dynamics, the electron is no longer neutral. The center of its atom controls the neutrality of that electron.

A structure evolves in one dimension if one electron of the gaseous atom executes interstate dynamics. There is a different nature of the binding energy. Before they execute confined interstate dynamics, the electrons of the gaseous atoms attain an orientation of ~ 5° along the north pole. In the dynamics of two electrons, a structure evolves in two dimensions. The electron dynamics of one cycle generate the binding energy. In the electron dynamics of the gaseous atoms, the binding energy mainly maintains the levitation nature of the force.

In the space format or atmosphere, an amalgamated atom and a targeted atom bind at the points of their generated energy. The atom binds from the upper side. The amalgamated or amalgamating atoms bind the targeted atom from the upper side. However, more work is required to establish a structural evolution in the atoms of gaseous elements.

### *3.2. Structural evolutions of semisolid element atoms*

Atoms of suitable semisolid elements amalgamate at the ground level, shown symbolically in Figure 2. To evolve the structure, electrons in the outer ring first maintain a neutral state, which is instantaneous. Figure 2 shows the distribution of forces. In the surface format, the forces of the four poles mainly contribute.

Atoms of semisolid elements grow to a suitable level. It is at the surface or ground level. Therefore, the forces exerted on the outer ring electrons of the semisolid atoms remain along all the poles. In a neutral state, electrons of suitable semisolid atoms maintain an orientation of almost zero degrees along the parallel lines passing through their centers.

Label (1) in Figure 2 shows the same orientations of the electrons belonging to all four quadrants. The presence of either two or four outer ring electrons is associated with conserved forces when the atoms reach the neutral state. The energy is already there to engage. Label (2) in Figure 2 denotes the ground points of the semisolid element atoms.

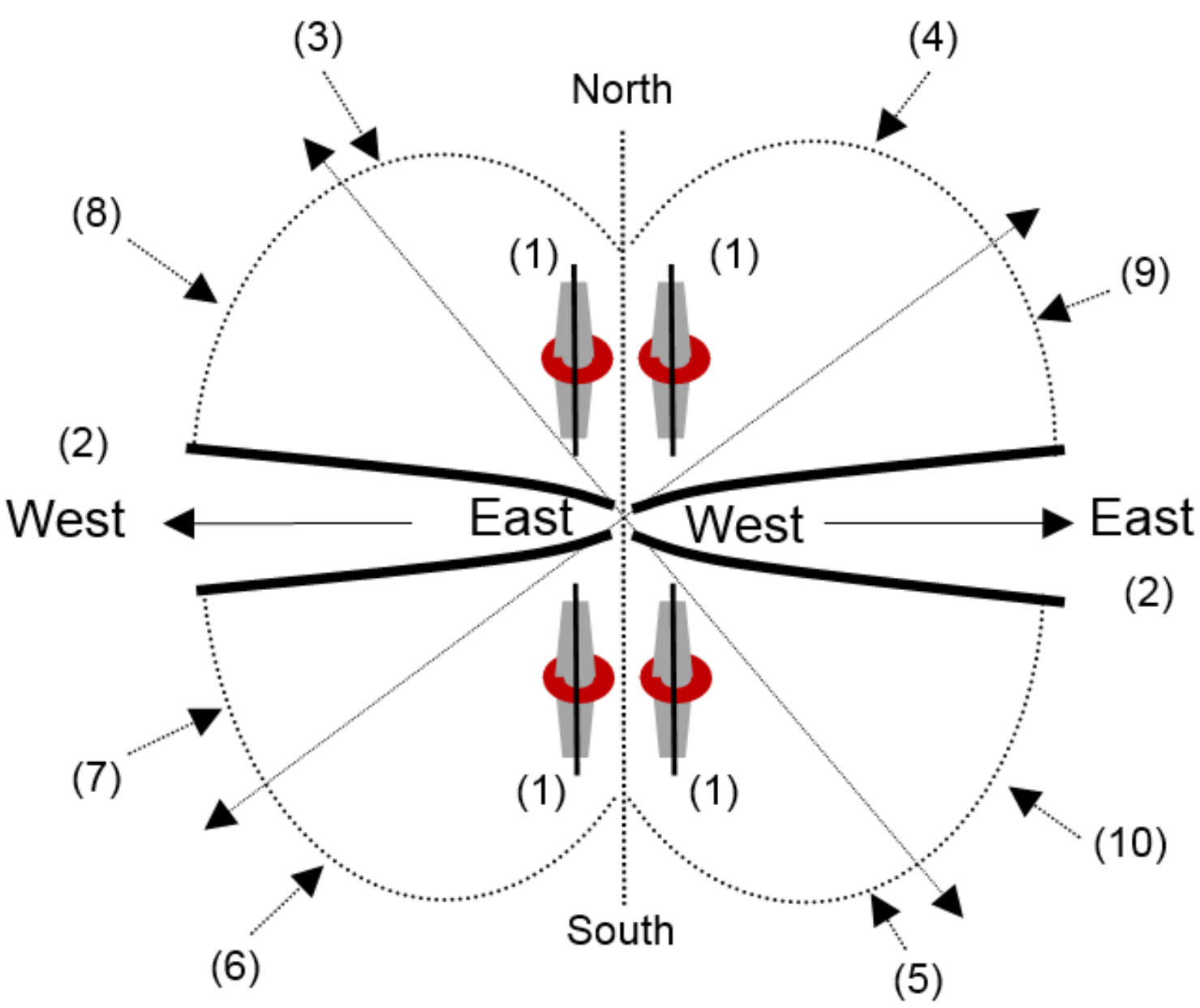

**Figure 2:** Exerted forces on the electrons of suitable semisolid element atoms to generate binding energy in one cycle of dynamics: (1) vertical lines drawn from the centers of left- and right-positioned electrons, (2) ground points of suitable semisolid

element atoms, (3) top left region posterior to the north pole, (4) top right region posterior to the north pole, (5) bottom right region posterior to the south pole, (6) bottom left region posterior to the south pole, (7) bottom west region, (8) top west region, (9) top east region and (10) bottom east region. (Sketch drawn in estimation)

Atoms of suitable semisolid elements exhibit the dynamics of two or four electrons. Under electron dynamics, this scheme keeps the atom in the equilibrium state, as discussed elsewhere [23]. Different labels (3 to 10) in Figure 2 indicate different regions of forces exerted on the left- and right-sided electrons of a semisolid atom. In Figure 2, a drawn vertical line with an electron has a 0° angle. Therefore, the forces of all four poles contribute.

The electrons keep a neutral behavior before their confined interstate dynamics, as discussed elsewhere [23]. To generate the binding energy, one cycle of electron dynamics is sufficient. The binding energy for a complete cycle of electron dynamics keeps the forces of all four poles. Figure 2 also shows the forces exerted on the electrons of suitable semisolid atoms. The binding of atoms is at the points of their energy. The binding of amalgamated atoms to the targeted atom is horizontal. However, carbon atoms form structures instead of evolving [25].

### *3.3. Structural evolution of solid element atoms*

The atoms of suitable solid elements amalgamate. In the amalgamation, atoms attain uniform dynamics. They keep the ground points below ground or at the surface level. Figure 3 shows this. The electronic orientation is from the south pole. Labels (1) and (2) in Figure 3 show the same orientations of the electrons. Label (3) in Figure 3 denotes an arc-shaped line. The arc-shaped line indicates the ground points of solid atoms in different elements.

Label (4) in Figure 3 indicates the east-west poles. In the structural evolution of solid atoms, a minor contribution of surface force remains. To evolve the structure, a suitable electron belonging to the outer ring first maintains the neutral state, which is instantaneous. The main exerted force on the electron remains along the south pole.

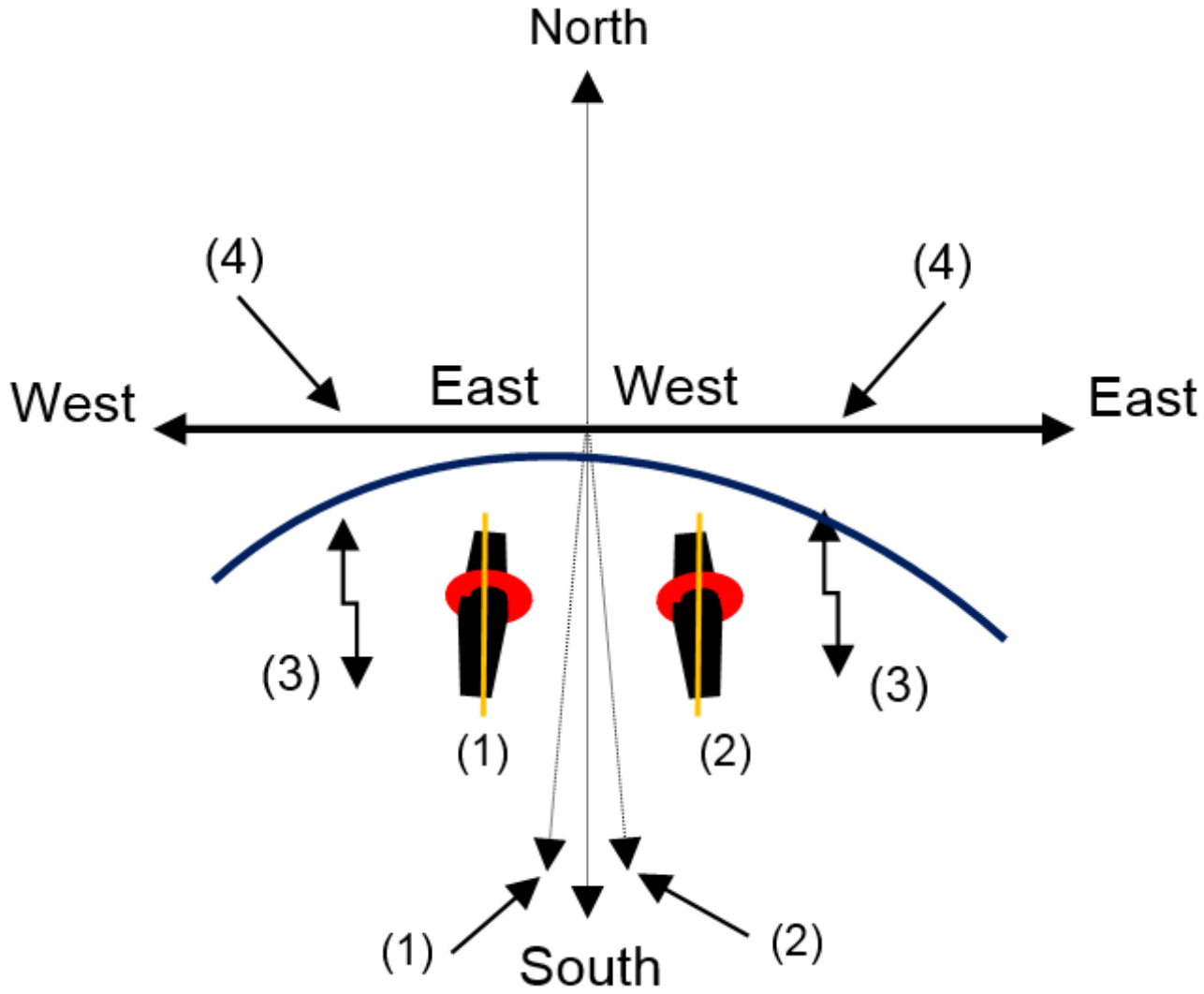

**Figure 3:** Exerted forces on the electrons of suitable solid element atoms to generate binding energy in one cycle of dynamics: (1) orientation of the left-sided electron ~ 5° left from the drawn vertical line, (2) orientation of the right-sided electron ~ 5° right from the drawn vertical line, (3) ground points of suitable solid element atoms and (4) ground or surface level. (Sketch drawn in estimation)

The surface force contribution, east or west, or both, is minor. These forces relate to a surface format. By adjusting the lattice, the suitable electron of the outer ring attains the neutral state. Therefore, the electron leaves the state to execute confined interstate dynamics. A force exertion is conservative. The energy is already present.

When a solid atom executes the dynamics of a suitable electron, the evolution of the structure is one-dimensional. The binding energy mainly reflects the gravitational nature of the force. A grounded format acts below the ground or surface level.

In the grounded format, an amalgamated atom and a targeted atom bind at the points of the generated energy. The binding of amalgamated atoms to the targeted atom is from the downward side. Atoms of suitable solid elements can evolve structures into different types.

### *3.4. Binding energy*

Path-independent but state-dependent forces are on the electrons of the outer ring in all three types of atoms – gaseous, semisolid, and solid. Engaged energy triggers the execution of electron dynamics, as discussed in sections 3.1, 3.2, and 3.3. Therefore, an atom under electron dynamics generates binding energy while involving conservative forces. The shape of the binding energy depends on the nature of the atomic element.

The binding energy of one cycle of electron dynamics in suitable atoms of the gaseous, semisolid, or solid elements has the shape of an interstate gap or distance. Figure 4 shows the binding energy in the tick, integral, Gaussian distribution, and letter *L* shapes. When the forces exerted on the electron are related to only two poles, the binding energy is like a tick symbol. Figure 4 (a) shows the shapes of this tick symbol.

When the conservative forces of only three poles act on the electron, the binding energy is an integral symbol. Figure 4 (b) shows the shapes of an integral symbol. There are also the four poles. Four poles introduce forces of all four main poles. The shape in Figure 4 (c) is the binding energy in this case.

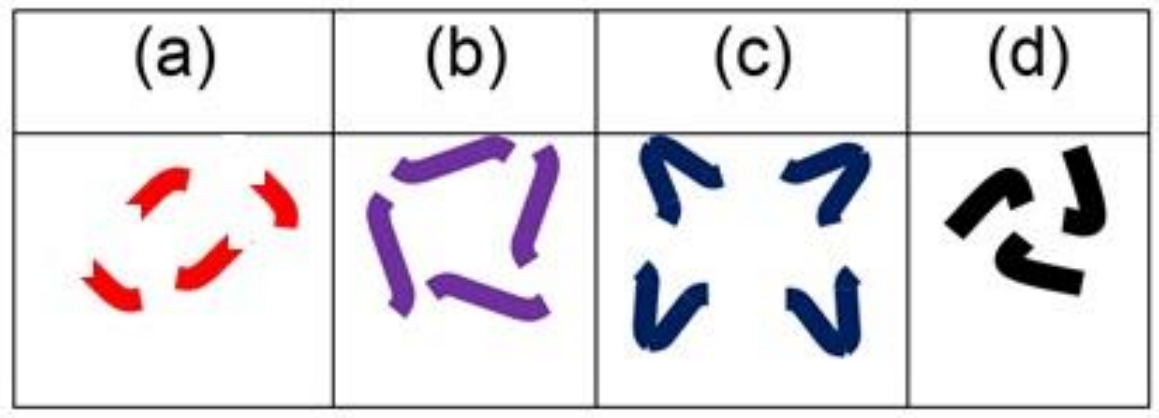

**Figure 4:** Binding energy in the (a) tick, (b) integral, and (c) Gaussian distribution of the turned ends, and (d) letter *L* shapes in atoms of different elements executing confined interstate electron dynamics

If three electrons (of the outer ring) execute the confined interstate dynamics, there is a three-dimensional structural evolution of the atoms. Here, the electrons revert without occupying (or touching) the nearby unfilled states. An electron (of the outer ring) traces an orientation at 120° in an atom. Three electrons of the outer ring simultaneously execute dynamics for one cycle in a slightly disturbing manner. The force exertion on the electron stops due to its disappearance. It is because of the appearance of the atomic pole itself. Therefore, an electron reverts without occupying the nearby unfilled state. The dynamics of the electron generate binding energy in the shape of the letter *L*. The binding energy in Figure 4 (d) shows the shape of the letter *L*.

The pole of an atom disappears at the last stage of the relevant exertion of force on the electron executing confined interstate dynamics. Thus, that electron cannot occupy the dedicated unfilled state. The tracing trajectory remains in letter *L*, rather than in an integral symbol. The remaining two electrons of the outer ring also face the same situation. Forces from two poles are exerted on the electron at each time of dynamics, as discussed elsewhere [23]. Figure 4 (a-d) shows the different shapes of the binding energy.

When there are many cycles of electron dynamics, the resulting energy relates to a photon energy, as discussed elsewhere [23]. The conditions of the space and grounded formats can deploy at the ground or surface level. Gaseous and solid atoms can also evolve structures at the ground or surface level. This approach can open many avenues of research.

### ***3.5. Three-dimensional structural evolution***

The evolution of three-dimensional structures is possible in suitable solid or gaseous atoms. The position of each filled state electron should be at a difference of 120° in the auxiliary outer ring of an atom. In each electron dynamics, there is a binding energy for one cycle. The resulting binding energy is in the shape of the letter *L*. For the evolution of structure in three dimensions, section 3.4 also keeps some details.

Figure 4 (d) shows the energy of one cycle of the electron dynamics. In the evolving three-dimensional structure, when attempting to transfer to the appropriate unfilled state, a suitably positioned electron cannot entirely cross the dedicated pole of the solid or gaseous atom. The retrieval of that electron occurs without contacting the nearby unfilled state or energy knot.

The electron faces a specific portion of the energy knot. The electron returns to the original state without occupying the unfilled state. The remaining two electrons of the outer ring also execute dynamics in the same manner. All three outer ring electrons attempted to cross the dedicated poles. Each positioned electron has an energy shape similar to the letter *L*. This binding energy results in a single cycle of electron dynamics.

Such evolved structures of solid atoms have a naturally finished surface. It is due to partial lateral and partial adjacent binding of the atoms.

In the binding energy, there is a minute level of the turning force. To some extent, the structures of solid atoms exhibit a ductile behavior. The atoms of suitable elements maintain the forces of the three poles. However, there is a need to trace the gaseous and solid atoms suitable for evolving three-dimensional structures.

### *3.6. Structural evolution of silicon atoms*

The forces of all four poles contribute to executing the confined interstate dynamics of the suitable electrons (of the outer ring) of a silicon atom. There is a conservative force for each electron. A silicon atom keeps the ground point at the ground or surface level. In the electron dynamics, the forces of a surface format mainly entail. The elements of levitational and gravitational forces are also there.

A study elsewhere [23] discussed the forces involved in turning the electron. A structure evolves in two dimensions when two electrons (of the outer ring) execute dynamics. Figure 5 (a) shows a two-dimensional structure. Both the amalgamated and targeted atoms bind at the binding energy points. The electron dynamics are for one cycle. Two amalgamated silicon atoms bind to the targeted atom at the points of generated energy.

Each amalgamated atom can also execute confined interstate dynamics of all four electrons of the outer ring. Hence, a four-dimensional structure evolves. Figure 5 (b) shows this structure. In this case, four amalgamated silicon atoms bind to the targeted silicon atom at the points of generated energy. More work is needed to obtain a complete picture.

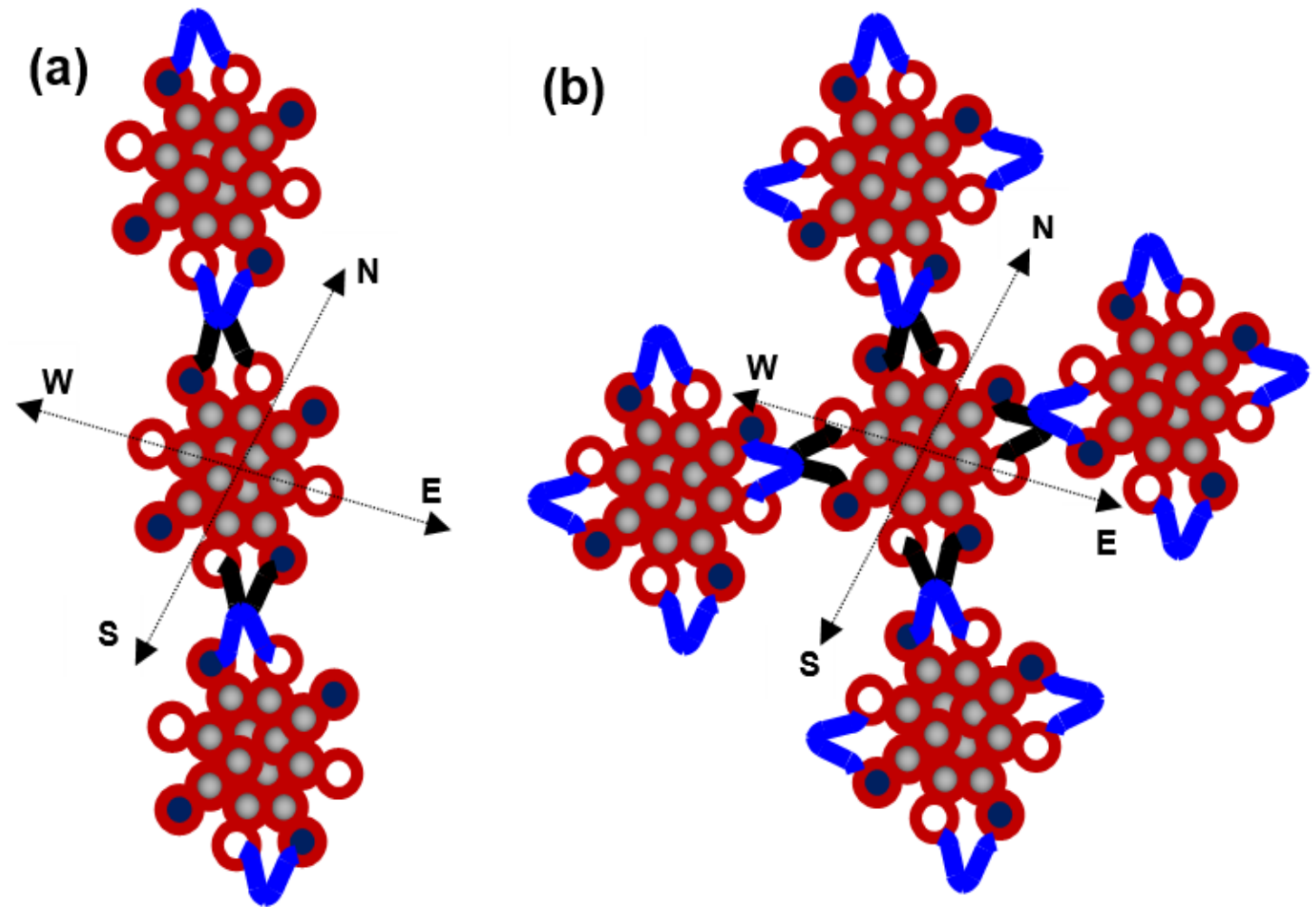

**Figure 5:** (a) Two-dimensional structure evolution in silicon atoms with confined interstate dynamics for two electrons and (b) four-dimensional structure evolution in silicon atoms with dynamics for four electrons. (Sketches drawn in estimation)

The generated binding energy of an atom under the electron dynamics in one quadrant would disturb the balance required for the equilibrium state of a silicon atom. Therefore, silicon atoms might not evolve structures in one dimension.

The dynamics of three electrons in three quadrants can disturb the balance of a silicon atom. Therefore, the evolution of structure in three dimensions is not feasible in silicon atoms. However, additional work is needed.

### *3.7. Binding of mono-layer shapes*

Two mono-layers bind in parallel. They involve the conservative force and engage the conservative energy in their binding. In the binding of mono-layers, atoms (in each mono-layer) align with the orientation of their electrons. Binding in mono-layers is due to the chemical activity of the conserved forces and energy. Each electron is associated with the conserved force and energy. Thus, each atom is associated with the conserved force and energy.

In the evolution process of their structure, a structure grows lengthwise and widthwise. The forces exerted at the electron level function from a distance. The energy at the electron level acts locally. When binding into a mono-layer shape, atoms preserve their force-energy, which is in chemical form at the electron level. There is a conserved force-energy for each mono-layer shape.

Lateral binding of the mono-layers involves force and engages energy in a dot-shaped bed. However, further studies are required to investigate the binding of different element atoms to single layers. A dot-shaped bed between mono-layers is due to the activities of force and energy. Initially, the activity of the force-energy is at the electron level. They are then at the atomic levels. Subsequently, they are at the mono-layer levels.

A chemical in nature, the force-energy trap between mono-layers, settling the binding atoms' expansion and contraction. There is also a need to study the dimensional changes. Due to a conserved force at the electron level, it can relate to a dot force. The dot force in the mono-layers of gaseous atoms can relate to their levitating nature. It relates to a gravitating nature in binding the mono-layers of solid atoms. There is a need to investigate these phenomena in detail.

### *3.8. Possibility of the surface plasmon phenomenon in a mono-layer tiny particle*

Several studies published in the literature cover the phenomenon of surface plasmons. As discussed in many studies, a surface plasmon phenomenon occurs in the developing tiny particles. There is a trapping of the photons that traveled along the air-solution interface.

As discussed in the published literature, the atoms of a tiny particle undergo collective oscillations in the surface plasmon phenomenon. However, the solid atoms can modify their electronic structure during amalgamation in any-sized cluster.

The published high-resolution microscopic images in the literature show different shapes of atoms of tiny particles. Those images indicate different modification behaviors of the atoms. Solid atoms at the electron level also deal with a surface force at the surface (or ground) level. Avoiding the exertion of that force is a grand challenge. The studies cited here also show atom-level modifications. The reference studies 1, 3, 4, 10, 11, 12, 14, 15, 22, and 28 discussed the atom-level modifications.

In the development process of a tiny particle, atoms of suitable elements do not retain their original shapes. As a result, the bound atoms should not oscillate collectively. The trapping of photons by the atomic lattices of a tiny particle is also not viable. When they interact with a suitable medium, photons can convert into pieces of heat energy (or energy bits) [23]. Photons are like X-rays, and they can also reflect [28].

A tiny particle should evolve its structure rather than develop. So, there is no atom-level modification. If the surface plasmon phenomenon is possible, an evolved tiny particle should retain a suitable number of atoms. The photons should travel with the required energy. The photons should travel in a specific direction relating to the tiny-sized particle with a mono-layer. Some other conditions for studying the surface plasmon phenomenon can occur. If the lattices of atoms, which belong to a tiny particle, oscillate collectively upon experiencing traveling photons. It is a wonder. The atoms of a mono-layer tiny particle retain the same shape during its evolution. Therefore, it is fascinating to see the execution of a surface plasmon phenomenon. It is also good to see the resulting outcomes of different tiny particles.

To qualify for the surface plasmon phenomenon, photons should travel under a suitable scheme. Photons should keep the required characteristics. The photons should travel for one instant parallel to the surface of a single-layer tiny particle. For the next instant, the photons travel from the opposite end. The process of traveling photons should remain intact for the required interval. The photons from the surface of a tiny particle should travel back and forth. Thus, a suitable tiny particle will experience a surface plasmon phenomenon.

Earlier studies related to atomic structure discussed shells, orbits, and the nucleus. A nucleus contained the protons and neutrons in the conventional studies. However, the earlier discussed electronic configurations do not allow the atoms to oscillate collectively. The phenomena of surface plasmons published in the literature are not in accordance with an actual atomic structure. Atoms keep energy knots clamping electrons and zeroth rings under a new insight.

There is a need to investigate the surface plasmon phenomenon in connection with binding patterns. Future research should also consider dimension changes in connection to a surface plasmon phenomenon. The surface plasmon phenomenon in the semisolid atoms is less favorable. However, the surface plasmon phenomenon in the context of metallic and other solid atoms is highly favorable.

There is a need to study the phenomena of surface plasmon polaritons and surface plasmon resonance. A surface plasmon polariton phenomenon is more intense than a surface plasmon resonance phenomenon. A study elsewhere [24] discussed atomic structure (in the periodic table elements) with a new insight. Thus, it is necessary to consider the possibility of two surface plasmon phenomena in the context of a new atomic structure.

### *3.9. General discussion*

The shape of the binding energy depends on the dynamics of an outer-ring electron. In atoms of suitable elements, electron dynamics occur in the gap (or distance) between the states. The trajectory of the electron in interstate dynamics is similar to the distance between the states. To attain neutrality, the electrons of gaseous atoms maintain an orientation of ~ 5° along the north pole, as discussed elsewhere [24].

To attain neutrality in solid atoms, the electrons maintain an orientation of ~ 5° along the south pole [24]. Again, electrons of semisolid atoms maintain an orientation almost parallel to the vertical lines passing through their centers [23]. Particles of different geometric shapes developed due to varying the electronic orientations of the atoms [28]. Entropy behaves consistently under plastically driven electronic states of the atoms, as discussed elsewhere [29].

Several studies have discussed the phenomena of surface plasmon polaritons and surface plasmon resonance. Only a few of them are cited here [30-55]. None of them addresses the actual phenomenon. Nevertheless, for the evolution of structures in suitable atoms, there is a need to develop a setup of space and grounded formats at the ground or surface level.

An evolved structure should behave differently from a developed structure in an X-ray analysis. Many avenues of researching materials can open up at this point. Different extracted ores are also processed to purify the atomic compositions. However, when atoms are dissociated from the precursor or ejected from the source target, they amalgamate or deposit to develop materials with different properties.

During the development of a material, atoms undergo different modifications. In developing a material, the force and energy are not conserved. In structural evolution, forces and energy become conserved. Atoms do not elongate for a structural evolution. Atoms also do not get a distorted shape. In the structural evolution, atoms do not undergo any deformation. Research on the structural evolution of suitable element atoms can open new fields of study.

## 4.0. Conclusion

When the atoms of suitable elements evolve structures, they occupy the zone of their original growth. Before binding, atoms amalgamate through uniformly attained dynamics. The electrons of the outer rings of suitable atoms deal with the conserved forces to engage the conserved energy within the interstate gaps.

The engagement of energy is due to the involvement of forces first. Before the dynamics of a suitable outer ring electron, an atom instantaneously attains neutrality for that electron. A one-dimensional structure in suitable elements evolves if one outer-ring electron of the atoms undergoes dynamics.

In the atoms of those elements, where three electrons of the outer ring can execute the dynamics, a structure evolves in three dimensions. When an electron deals with the forces of two poles, the binding energy of each electron maintains the shape of a tick symbol. If the two outer-ring electrons of the atoms execute the interstate dynamics, the structure evolves in two dimensions. When an electron deals with the forces of three poles, the binding energy of each electron maintains the shape of an integral symbol. However, at one time, only two forces are exerted [23].

When the electron dynamics deal with the disturbance at the end, despite involving the forces from three poles, the binding energy has the shape of the letter *L*. In this case, the three outer-ring electrons of the atoms execute the interstate dynamics. The structural evolution is in three dimensions. In the outer ring of the atom, there is a ~ 120° angle between the electrons.

The structure of suitable semisolid atoms is four-dimensional. For this, all four electrons of the outer ring execute dynamics. Here, the shape of the generated binding energy is similar to the Gaussian distribution of the turned ends. Atoms of suitable semisolid, solid, and gaseous elements bind side-to-side, laterally from the downsides, and laterally from the upper sides, respectively. The electron dynamics of one cycle generate binding energy.

A mono-layer binds to another mono-layer from the downward side. It occurs in the atoms of the solid elements. The binding of mono-layers of suitable gaseous element atoms is from the upward side. Mono-layers of bound atoms bind due to the chemical activity of conserved forces and energy. A surface plasmon phenomenon exists in the evolved structure of a single-layer tiny particle, as the electronic configuration of all the atoms remains alive. During the development of a single-layer tiny particle, atoms do not maintain the same electronic orientation as in their original state. Therefore, they cannot execute different surface plasmon phenomena.

A tiny particle of the mono-layer can oscillate its atoms collectively under suitable conditions. How do the atoms in various elements attain their neutral states? How can a cluster of a few atoms oscillate collectively under a suitable arrangement? Such questions leave an intriguing note. It is challenging for chemists to quantify the rates of various activities and their consolidation. The DFT and MD simulations can investigate the correlation between electron dynamics and binding energy. These would strengthen the theoretical aspect. The experimental investigations are also needed to validate this study.

## Statements & Declarations

**Funding:**

No funding was received for conducting this study.

**Competing Interests:**

The author has no relevant financial or non-financial interests to disclose.

**Author Contributions:**

Not applicable. This work was prepared as a solo author.

**Data availability statement:**

This manuscript does not contain any associated data.

## Bibliographic detail:

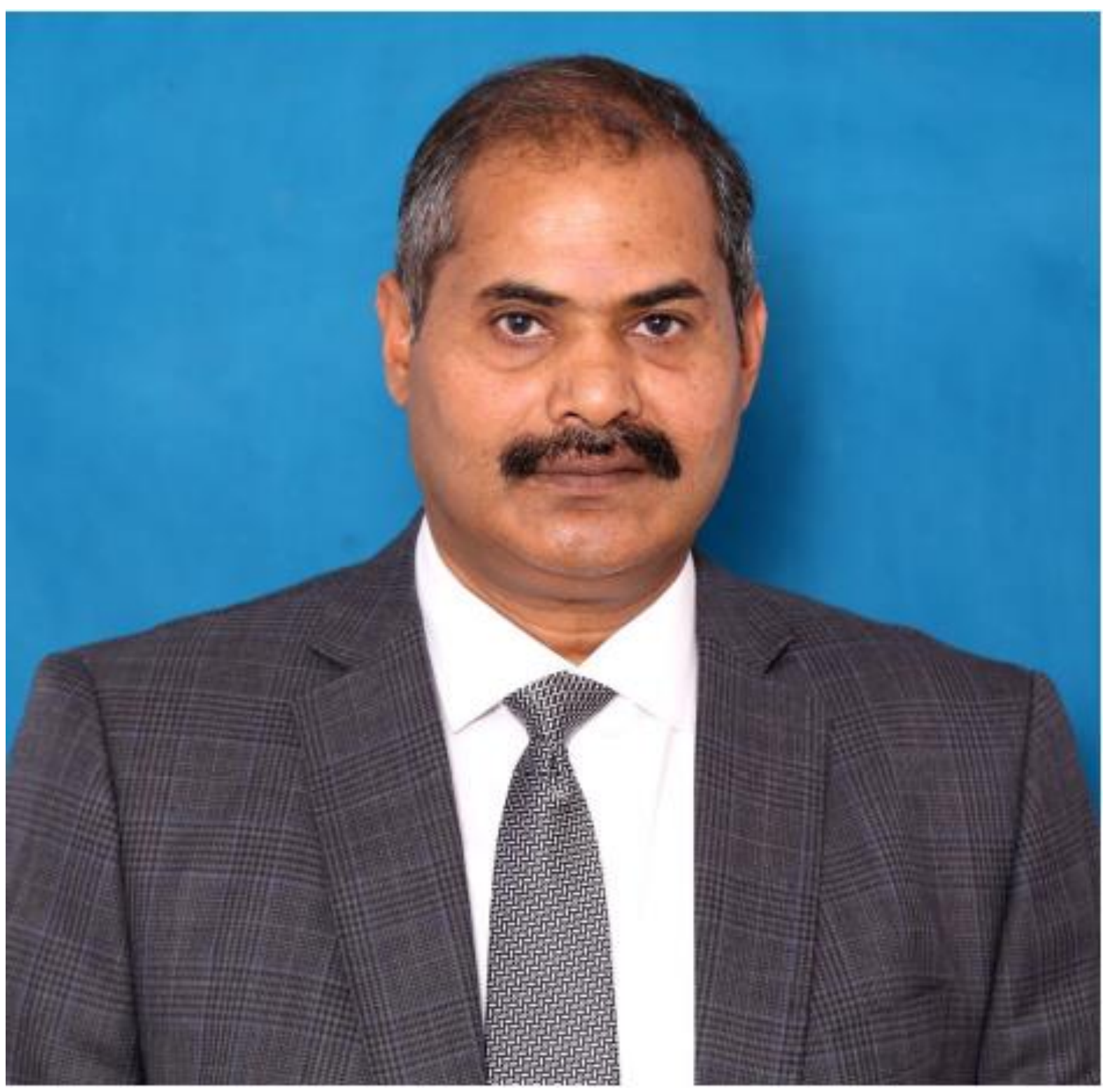

In 1996, **Mubarak Ali** earned a B.Sc. degree in Physics and Mathematics. The University of the Punjab awarded him a degree. His M.Sc. degree in Materials Science in 1998. Bahauddin Zakariya University Multan awarded him a master's degree with distinction. He completed his thesis at Quaid-i-Azam University Islamabad. He gained a PhD in Mechanical Engineering from the Universiti Teknologi Malaysia under the award of the Malaysian Technical Cooperation Programme (MTCP;2004-07) and a postdoc in advanced surface technologies at Istanbul Technical University under the foreign fellowship of The Scientific and Technological Research Council of Turkey (TÜBİTAK, 2010). Dr Mubarak completed another postdoc in nanotechnology at Tamkang University Taipei, 2013-2014, sponsored by the National Science Council, now the Ministry of Science and Technology, Taiwan. He remained working as an Assistant Professor on the tenure track at COMSATS University Islamabad from May 2008 to June 2018, previously known as the COMSATS Institute of Information Technology. His new position is in process. He also worked as an assistant director and deputy director at M/o Science & Technology, Pakistan Council of Renewable Energy Technologies, Islamabad, from January 2000 to May 2008. The Institute for Materials Research at Tohoku University Japan invited Dr. Mubarak to deliver a scientific talk. His scientific research has been a part of many conferences organized by renowned universities in many countries. His core areas of research include materials science, physics, surface and coating technology, carbon-based materials, materials engineering, materials chemistry, physical chemistry, sustainability, energy science, and nanotechnology. He also won a merit scholarship for PhD studies from the Higher Education Commission, Government of Pakistan. However, he did not obtain this opportunity. He earned a diploma (in English) and a certificate (in the Japanese language) in 2000 and 2001, respectively, part-time from the National University of Modern Languages, Islamabad. He is the author of several articles. Please refer to the link https://www.researchgate.net/profile/Mubarak_Ali5 and the link https://scholar.google.com.pk/citations?hl=en&user=UYjvhDwAAAAJ